\documentclass[a4paper,11pt]{article}
\usepackage{pos}

\title{Single- and double-heavy Hadronic Molecules}
\author*[a]{C. Hanhart}
\affiliation[a]{Institute for Advanced Simulation (IAS-4), Forschungszentrum J\"ulich, J\"ulich, Germany
  }

\emailAdd{c.hanhart@fz-juelich.de}
\abstract{In this presentation the notion of hadronic molecules is reviewed and it is argued that some of the
 enigmatic single and double heavy mesons, namely the lowest lying positive parity open charm states, the $T_{cc}(3875)^+$
 and
 the $\chi_{c1}(3872)$ aka $X(3872)$, that do not fit into the conventional quark--anti-quark scheme in fact
 qualify as hadronic molecules. For the single heavy states we show that an alternative explanation as diquark--anti-diquark 
 structure is at odds with either phenomenology or lattice data.
 For the $X(3872)$ we discuss also the claimed isovector partner state, whose properties
  would provide additional strong support for a molecular structure
 of the $J^{PC}=1^{++}$ states near the $D\bar D^{*}$ thresholds. Its existence could be confirmed
 by, e.g., a high statistics measurement of the $J/\psi\pi^+\pi^-$ lineshape from $B^0\to K^0  J/\psi\pi^+\pi^-$.
 }
 
\FullConference{The 21st International Conference on Hadron Spectroscopy and Structure (HADRON2025)\\
 27 - 31 March, 2025\\
Osaka University, Japan\\}

\begin{document}
\maketitle

\section{Introduction}

Since the beginning of this century hadron spectroscopy regained focus of the physics community
triggered by a series of spectacular observations in the sector of single and double heavy hadrons---for reviews, putting emphasis on different aspects, see Refs.~\cite{Hosaka:2016pey,Esposito:2016noz,Guo:2017jvc,Olsen:2017bmm,Karliner:2017qhf,Brambilla:2019esw, Yang:2020atz,Chen:2022asf,Meng:2022ozq,Hanhart:2025bun}.
Many of those states are located close to $S$-wave thresholds of pairs of long lived or, with respect to the strong interaction, stable particles
and thus appear to be good candidates for hadronic molecules or, more generally, two-hadron (or even few-hadron) states. A state qualifies as hadronic molecule, if all its properties are consistent
with its structure being composed of two (or more) hadrons---a well known example of such a state is the deuteron. Since the hadron pair can go on the 
mass shell above the threshold, generating an imaginary part
in the scattering amplitude that is absent below threshold, there is a branch point at the threshold  (for a detailed discussion
of the analytic properties of scattering amplitudes see Refs.~\cite{resonancesPDG:2024cfk,Mai:2022eur}), which leaves an imprint in properly chosen observables. 
In contrast to this, diquarks carry a color charge and therefore they are confined within the hadron. This implies that there is
no prominent cut in the amplitude, if mesons were dominated by diquark--anti-diquark structures and the above mentioned
observables are predicted to be different.

In this writeup I present the overwhelming evidence, why the lowest lying positive parity open charm states
as well as the $\chi_{c1}(3872)$ aka $X(3872)$ should be classified
as hadronic molecules. The analyses performed employ chiral effective field theory as well as, for the former
case, lattice QCD. Moreover, the Weinberg criterion and extensions thereof are used to identify the nature
of the states. Accordingly, the write-up is structured as follows: In the next section the Weinberg criterion is 
introduced. Then we present a short description of chiral perturbation
theory (ChPT) and the unitarisation thereof (lattice QCD was presented in depth in other presentations at the
conference), before we turn to the discussion of the examples. The presentation closes with a short summary.

\section{The Weinberg criterion}

Before discussing the implications of a molecular or more general few-hadron structure of certain families of
states in some depth, we first need a criterion to identify those states from observable quantities.
Thus a criterion was provided by Weinberg already in the 1965~\cite{Weinberg:1965zz} (inelastic channels were later included in Ref.~\cite{Baru:2003qq}). 
We here give a very short summary of the key result. For more details see Refs.~\cite{Guo:2017jvc,Hanhart:2025bun}.

For a single channel system the scattering amplitude for two-hadrons in the presence of a near threshold bound-state
pole
at $E=-E_B$, where $E$ denotes the kinetic energy of the hadron pair,
can be expressed as~\cite{Guo:2017jvc}
\begin{equation}
T_{\rm NR}=\frac{g_0^2}{E+E_B+g_0^2\mu/(2\pi)(ip+\gamma)} \ ,
\label{eq:TNRpole}
\end{equation}
 in the region very close to threshold. In Eq.~(\ref{eq:TNRpole}) $\gamma=\sqrt{2\mu E_B}$ denotes the binding momentum
with $\mu = m_1 m_2/(m_1+m_2)$ for the reduced mass. The term proportional to $\gamma$ needs to be 
introduced to cancel the analytic continuation of the $ip$ term a $E= -E_B$\footnote{A bound state is a pole on the physical sheet, where for 
negative energies $p\to i\sqrt{2\mu|E|}$; for the unphysical sheet we have for
the analytic continuation $p\to -i\sqrt{2\mu|E|}$.}---for virtual states this term needs to be changed to
$-\gamma$~\cite{Guo:2017jvc}.
With this,  the finding of Ref.~\cite{Weinberg:1965zz} can be condensed into
 \begin{equation}
g_0^2 = \frac{2\pi\gamma}{\mu^2}\left(\frac{1}{\lambda^2}-1+\mathcal{O}\left(\frac{\gamma}{\beta}\right)
\right) \ ,
\label{eq:g0oflambda}
\end{equation}
where $\beta$ denotes the intrinsic momentum scale of the vertex functions, which may be identified
with the mass of the lightest exchange particle allowed. 
The quantity $\lambda^2$ is the probability to find the compact component inside the hadronic
state---accordingly we have $g_0^2\to 0$ for a purely compact state and $g_0^2\to \infty$ for a purely molecular
state. Thus Eq.~(\ref{eq:g0oflambda}) provides a relation between the coupling of a bound state to the continuum
and the structure of the state. To relate this to observables we may employ the 
effective range expansion (ERE), defined via
\begin{equation}
T_{\rm NR}(E)=-\frac{2\pi}{\mu}\frac{1}{1/a+(r/2)p^2-ip} \ ,
\label{eq:ERE}
\end{equation}
with $a$ and $r$ for the scattering length and the effective range, respectively\footnote{We use the particle
physics sign convention for the scattering length.}, that can be determined from fits to data.
Matching Eq.~(\ref{eq:ERE})
to Eq.~(\ref{eq:TNRpole}) gives
\begin{align}
\frac{1}{a} &=-\frac{2\pi E_B}{\mu g_0^2}-\gamma & \Longrightarrow & \qquad \qquad {a=-2\frac{1-\lambda^2}{2-\lambda^2}\left(\frac{1}{\gamma}\right)  
+{\cal O}\left(\frac{1}{\beta}\right)} \ , 
\label{eq:arelations}
\\
r &= -\frac{2\pi}{g_0^2\mu^2} & \Longrightarrow &  \qquad  \qquad  {r= -\frac{\lambda^2}{1-\lambda^2}\left(\frac{1}{\gamma}\right)  +{\cal O}\left(\frac{1}{\beta}\right)} \ ,
\label{eq:rrelations}
\end{align}
using Eq.~(\ref{eq:g0oflambda}).
Thus, for a pure molecule ($\lambda^2=0$) we have $a=-1/\gamma+{\cal O}\left({1}/{\beta}\right)$ and
$r={\cal O}\left({1}/{\beta}\right)$
 and for a purely compact structure ($\lambda^2=1$)  
$a={\cal O}\left({1}/{\beta}\right)$ and $r\to -\infty$. In single channel scattering the corrections to $r$
are larger than zero~\cite{Bethe:1949yr,Esposito:2021vhu} such that
 a positive
effective range is an unambiguous signature of a purely molecular state~\cite{Weinberg:1965zz,Esposito:2021vhu,Baru:2021ldu,Li:2021cue,Albaladejo:2022sux}.
In general, however, not even a sizeable   negative effective effective range
provides a unique signature of a compact structure, since coupled channel effects also induce 
a negative contribution to the effective range that needs to be corrected for~\cite{Baru:2021ldu} before
applying the Weinberg criterion. Note that one should not use $a$ and $r$ as input to extract $\lambda^2$~\cite{Song:2022yvz}, but
more use the relations of Eqs.~(\ref{eq:arelations}) and (\ref{eq:rrelations})  to check if, within errors, the properties of a given system are consistent with
e.g. a pure molecule ($\lambda^2=0$).
For generalizations of Weinberg's
criterion also to virtual states and resonances see Refs.~\cite{Baru:2003qq,Matuschek:2020gqe,Bruns:2019xgo,Oller:2017alp,Kang:2016jxw,Sekihara:2016xnq,Kamiya:2016oao,Guo:2015daa,Sekihara:2014kya,Aceti:2012dd,Gamermann:2009uq,Kinugawa:2024kwb,Liu:2024uxn}.

\section{Chiral effective field theory}

\begin{figure}
\parbox{6cm}{\includegraphics[width=\linewidth]{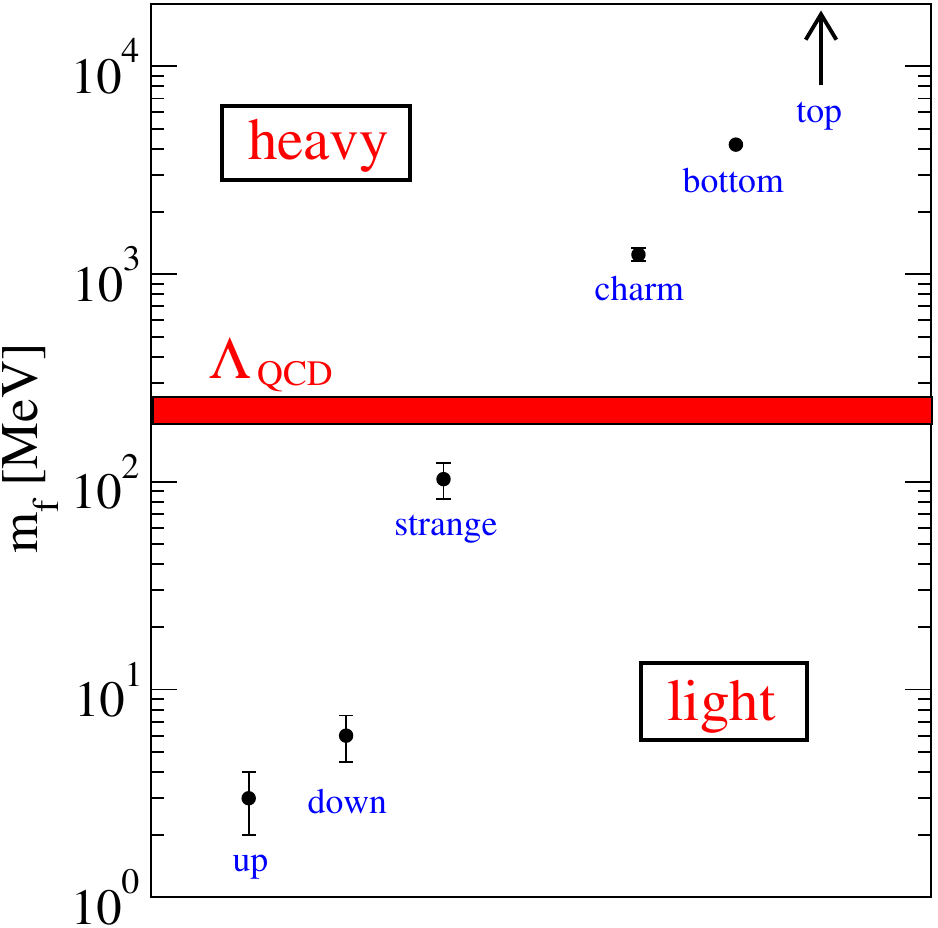}}
\parbox{9cm}{
\caption{Quark masses in comparison to the QCD mass scale $\Lambda_{QCD}$.
The masses of up, down and strange are the $\overline{MS}$ masses at a renormalisation scale
of 2~GeV and the heavy masses are evaluated at their own mass as scale~\cite{ParticleDataGroup:2024cfk}.
QCD allows for two systematic expansions, one around vanishing quark masses with corrections
of order $(m_q/\Lambda_{\rm QCD})$, applicable to up, down and eventually strange quarks,
 that can be treated using ChPT, and one around infinitely
large quark masses with corrections of order $(\Lambda_{\rm QCD}/m_Q)$, applicable to
charm, bottom and top quarks, that can be treated, e.g.,
using heavy quark effective field theory, (p)NRQCD or BOEFT. There is no systematic way to treat e.g. the light
quarks and the heavy quarks on equal footing.
 \label{fig:quarkmasses}
}}
\end{figure}

Chiral perturbation theory (ChPT) is a well established effective field theory for the Standard Model 
at low energies---for a recent
review see~\cite{EFTbook}.
It not only allows for the evaluation of hadronic amplitudes with controllable accuracy, but also provides
a basis for a study of hadronic resonances---for the alternative approach via the recently developed BOEFT see Ref.~\cite{Berwein:2024ztx}.
 ChPT exploits systematically the implications of the Goldstone theorem
related to the spontaneous symmetry breaking of the chiral symmetry group of massless QCD,
$SU(3)_L~{\times}~SU(3)_R~{\to}~SU(3)_V$, namely that 
the masses and the interactions of the pseudoscalar ground states, below generically
called $\phi$, that can be identified with the Goldstone bosons, vanish
for vanishing up-, down- and strange-quark masses  and hadron momenta. Then ChPT allows one to include systematically corrections for 
the finite quark masses (as an expansion in $(m_q/\Lambda_{\rm QCD})$ that in hadronic terms
translates to $(M_{\rm ps}/\Lambda_\chi)$, where $M_{\rm ps}$ denotes the masses of
the pseudo-scalar mesons that are now pseudo-Goldstone bosons and $\Lambda_\chi\sim 1$~GeV the typical hadronic mass scale)
 as well as the finite hadron momenta (as an expansion in $(p/\Lambda_\chi$)). While the theory evolving in this way is formally non-renormalisable
and characterised by infinitely many terms, it gets predictive once a proper power counting is introduced that
at the same time allows for a renormalisation order-by-order. The two-pion system is one of the best studied system in this context.
For a review see Ref.~\cite{Colangelo:2001df}.

The scheme can be extended to include heavy sources like nucleons~\cite{Gasser:1987rb,Bernard:1995dp} or hadrons
that contain heavy quarks~\cite{Manohar:2000dt}. It then turns out that, e.g., the amplitude for scattering of the
pseudoscalar ground states off heavy sources for a given set of flavor quantum numbers $a$
 at leading order in $S$-wave in the cms can be expressed as
\begin{equation}
{\mathcal A}_a=C_a E_{\phi\, a}/F_0^2 \qquad (F_0=\pi \ \mbox{decay constant}) \ ,
\label{eq:LOhl}
\end{equation}
where the parameters $C_a$ can be determined from Group Theory and $F_0$ from the weak
decay of charged pions~\footnote{Formally $F_0$ is the pion decay constant in the chiral limit,
but differences to the physical value or of higher order in the expansion}. 
Here $E_\phi$ denotes the energy of the pseudo-scalar mesons.
Thus, there is no
free parameter at leading order, which is known as the Weinberg-Tomozawa theorem~\cite{Weinberg:1966kf,Tomozawa:1966jm}.
Especially, if the chiral expansion converges, the sign of the leading interaction, which is fixed from the
start, decides, if the interaction is attractive or repulsive. This observation is crucial for the discussion of
heavy-light systems in the next section. For the scattering of pseudo-scalars off heavy sources, at 
next-to-leading order additional tree level contributions enter
with a few parameters to be determined from fits to experiment or to lattice data. Loops enter one order higher.

As a direct consequence of the Goldstone theorem, Eq.~(\ref{eq:LOhl}) contains two features crucial
for the following discussion: The leading interaction grows quickly as the energy is increased and the 
interaction of kaons and eta mesons is stronger than that of pions. The former  has the consequence
that the amplitude hits the unitarity limit rather quickly. Then the range of applicability of the theory needs to
be extended by a proper unitarisation~\cite{Oller:1997ti,Oller:1998hw} that can be understood in
the spirit of a dispersive representation of the inverse scattering amplitude with the subtraction
terms fixed by ChPT~\cite{Dobado:1996ps,Oller:2000fj}. 
The respective studies of  heavy-light systems of interest below were initiated in Refs.~\cite{Kolomeitsev:2003ac,Guo:2006fu}.
As a welcome side effect this at the same time
enables the equations to dynamically generate poles in the scattering amplitude making the
formalism ideal for the study of hadronic molecules. The latter provides the reason, why in the
multiplets of two-hadron states those with kaons and eta mesons show significantly 
more attraction that those with pions---for a comprehensive discussion of the implications
for systems with kaons and heavy mesons see Ref.~\cite{Guo:2011dd}.

The situation is somewhat different for the scattering of two heavy sources: Here the leading-order interaction,
that now 
contains a free parameter (the strength of a contact term with 4 external legs) and, in the formulation
we shall employ below, the one-pion exchange, is energy independent. However, still a resummation is
necessary, since the intermediate propagator of the two heavy fields that comes in in this
way gets enhanced by a factor $\pi \mu/p$---it is this enhancement that can be understood as the 
origin of the existence of nuclei~\cite{Weinberg:1991um} (for a comprehensive review
see Ref.~\cite{Epelbaum:2008ga}). Especially for the study of doubly-heavy systems 
in the charm sector it is important to note that the one-pion exchange contribution introduces
a three-body cut at the nominal $D\bar D^*$ threshold. Already for pion masses only slightly above the physical value it also
generates a left-hand cut that then starts very close to the physical axis and thus needs to be considered
 for the chiral extrapolation of the pertinent systems~\cite{Du:2023hlu,Meng:2023bmz,Collins:2024sfi,Hansen:2024ffk}.
  At next-to-leading order pion mass and $p^2$ dependent contact terms enter as well
  as pion loops---note that it was shown by an explicit calculation that the
  latter can be absorbed into the counter terms~\cite{Chacko:2024cax}.

\section{Singly heavy states: The family of positive parity open charm states}

\begin{figure}
\parbox{11cm}{\includegraphics[width=\linewidth]{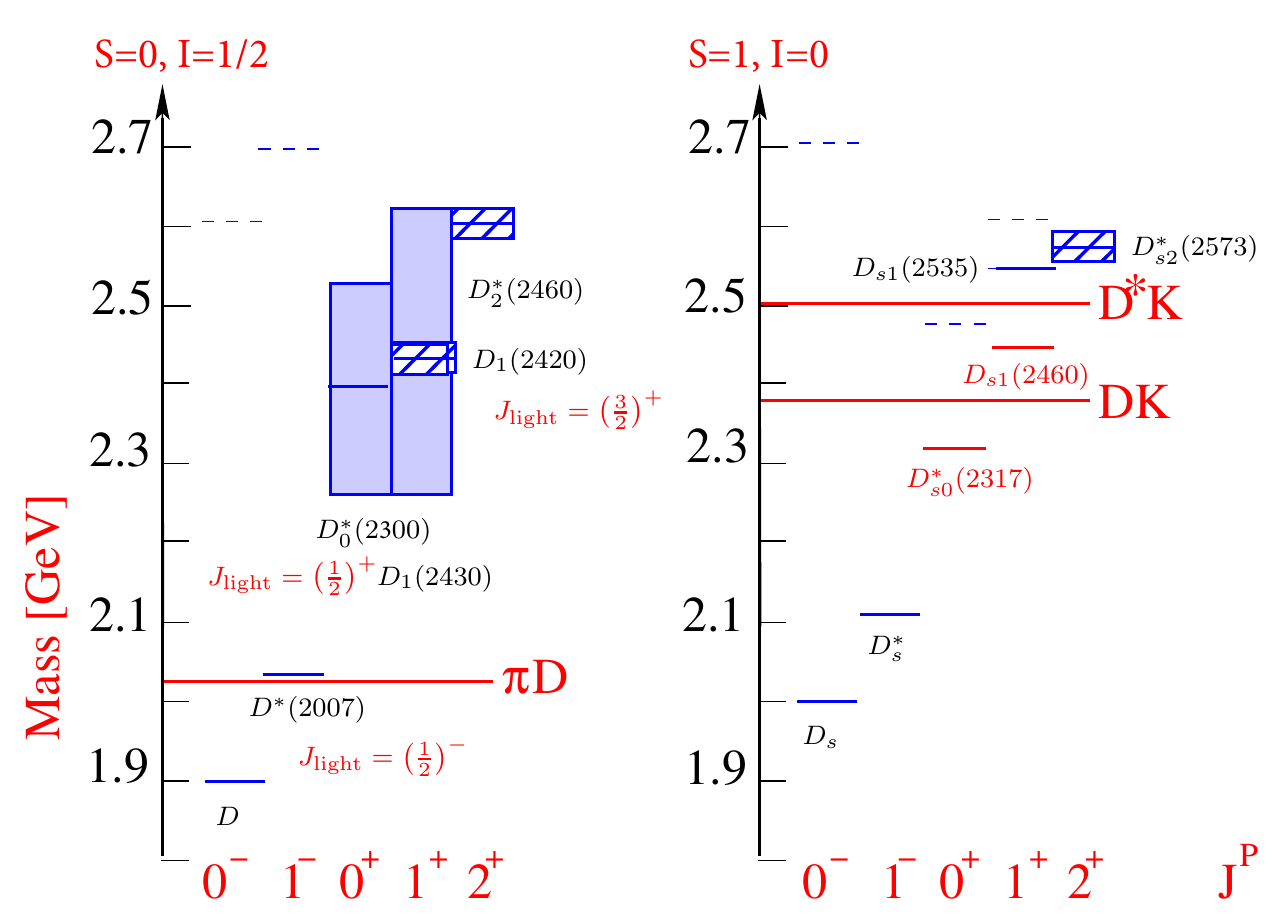}}
\parbox{4cm}{
\caption{Spectrum of the lightest positive parity non-strange (left) and strange
open charm states. Blue solid (dashed) lines show states predicted by the representative quark model
of Ref.~\cite{DiPierro:2001dwf}
and (not) confirmed by experiment and red solid lines show states found experimentally with properties at
odds with the quark model. The boxes indicate the widths of the states. \label{fig:Dspectrum}
}}
\end{figure}

With the discovery of $D_{s0}^*(2317)$ in 2003~\cite{BaBar:2003oey} and of its non-strange partner the $D_{0}^*(2300)$~\cite{Belle:2003nsh} shortly afterwards
began a quest for the nature of these states. The driving questions, mostly illustrated in Fig.~\ref{fig:Dspectrum}, can be summarised as
\begin{enumerate}
\item[$(i)$] Why is 
$M(D_{s1}){-}M(D_{s0}^*)\simeq  M(D^*){-}M(D)$?
\item[$(ii)$] Why are
$M(D_{s1}) \& M(D_{s0}^*)$ so much lighter than expected from the quark models (see, e.g., Ref.~\cite{DiPierro:2001dwf}) that assume a $c\bar q$ structure?
\item[$(iii)$]  Why are
$M(D_0^*)\simeq M(D_{s0}^*)$ and $M(D_1)\simeq M(D_{s1})$ 
although usually the presence of strange quarks enhance hadron masses by at least 100~MeV?
\item[$(iv)$] 
Why do unitarised ChPT and lattice analyses provide masses for the lightest $D_0^*$ state
so much smaller than the mass extracted from experiment?
\end{enumerate}

As we will demonstrate, not only provides the hadronic molecule picture---or, for resonances like
the $D_0^*$ one should better say two-hadron structures---a natural solution to all
those questions, but also are  recent lattice data inconsistent with both a $c\bar q$ as well as a diquark--anti-diquark structure. 

\begin{figure}
\parbox{7cm}{\includegraphics[width=\linewidth]{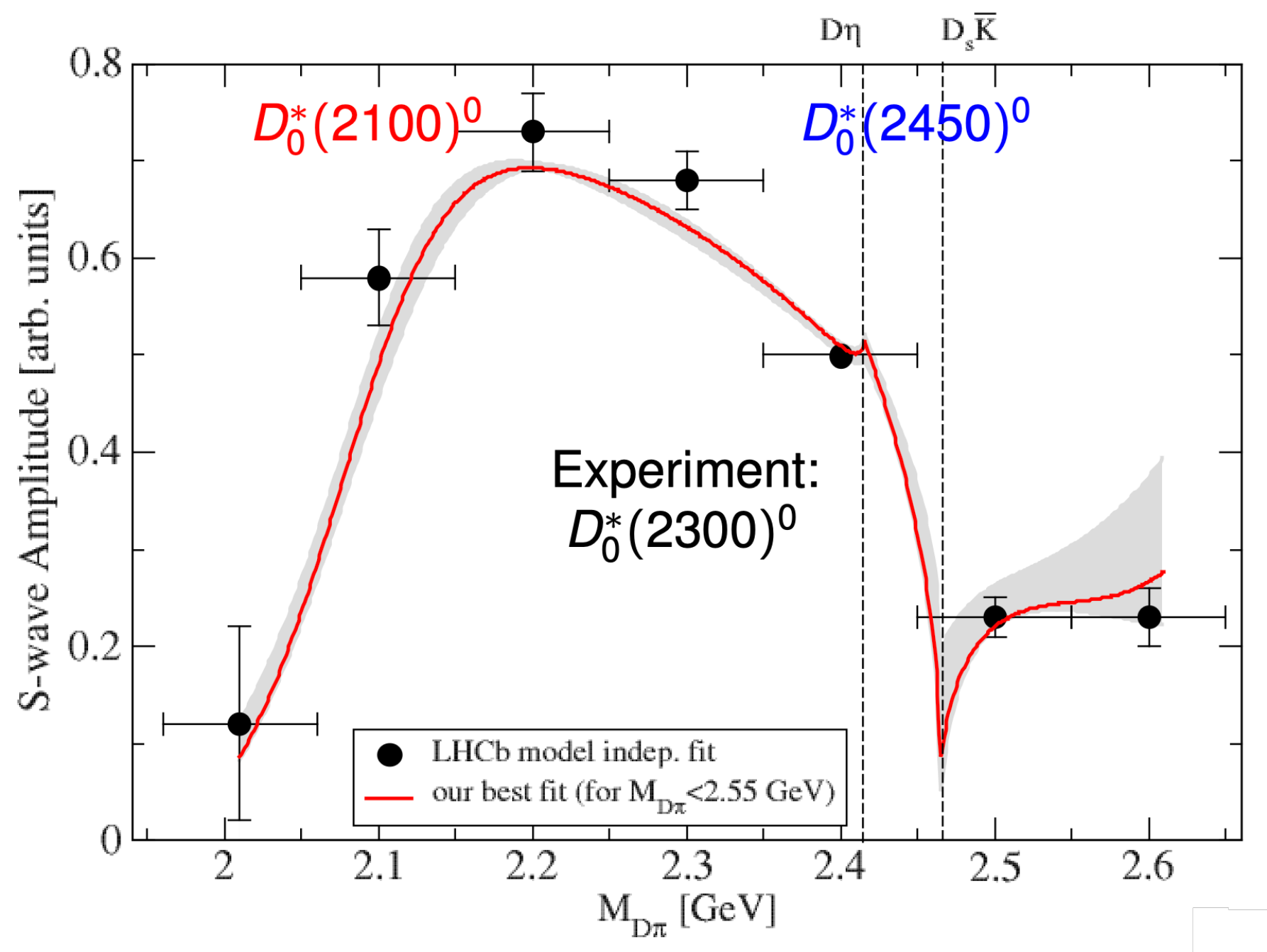}}
\parbox{8cm}{\caption{Fit result of Ref.~\cite{Du:2017zvv} employing the scattering amplitude of Ref.~\cite{Liu:2012zya}
to the experimental data of Ref.~\cite{LHCb:2016lxy} for the $\pi^- D^+$ $S$-wave
amplitude extracted from $B^-\to D^+\pi^-\pi^-$. The strong rise at low $M_{D\pi}$ is driven
by the pole at 2.1~GeV and the pronounced cusp near the higher thresholds (indicated by the
dashed lines)
is a signature of the pole at 2.45~GeV on a hidden sheet. However, fitting a single symmetric Breit-Wigner
function to those data naturally returns a mass of 2.3 GeV.
 \label{fig:Dpispectrum}
}}
\end{figure}

If indeed the lightest positive parity open-charm states owe their existence to non-pertrubative hadron-hadron dynamics,
it is natural that their masses are not dictated by the quark model (item $(ii)$), but get correlated with the pertinent 
thresholds. Moreover, in this picture the binding energy of kaons and $D$-mesons should equal that of
kaons and $D^*$-mesons up to spin symmetry violating effects that scale as $(\Lambda_{\rm QCD}/m_c)\approx 10$\%,
solving item $(i)$.
To address items $(iii)$ and $(iv)$ one needs to have a closer look at the driving interaction given in Eq.~(\ref{eq:LOhl}).
First of all one understands immediately, why the $KD^{(*)}$ systems develop bound states, while
the $\pi D^{(*)}$ systems only resonances---for a detailed discussion of the pole trajectories of 
systems containing pseudo-Goldstone bosons as the quark masses get changed, see
Refs.~\cite{Hanhart:2008mx,Guo:2009ct,Hanhart:2014ssa}. Moreover,
from the Group Theory point of view the scattering of the pseudo-scalar mesons off $D$-mesons calls for
the decomposition 
\begin{equation}
    [\mathbf{\bar 3}]\otimes [\mathbf{8}] = [\mathbf{\bar 3}] \oplus[\mathbf{6}]\oplus
[{\mathbf{\overline{15}}}] \ .
\label{eq:mol_decomp}
\end{equation}
The emerging strength parameters $C_a$, see Eq.~(\ref{eq:LOhl}), turn out such that the $[\mathbf{\bar 3}]$
feels the strongest attraction, a somewhat weaker one is found for $[\mathbf{6}]$
and the $[{\mathbf{\overline{15}}}]$ is repulsive~\cite{Kolomeitsev:2003ac}.
Employing a chiral-unitary amplitude~\cite{Liu:2012zya}, whose parameters were fixed by fits to lattice data
for the pion mass dependence of certain
heavy light scattering lengths, the authors of~\cite{Albaladejo:2016lbb} demonstrated
that not only the recent lattice data for $\pi D$ scattering above threshold of Ref.~\cite{Moir:2016srx} at 391~MeV pion mass
are fully consistent with the amplitude
determined earlier at threshold, but also contains two poles, one at 2264~GeV (belonging predominantly to
the $[\mathbf{\bar 3}]$) and one at 2468~MeV (belonging predominantly to the $[\mathbf{6}]$)\footnote{The lattice
group claimed only a single pole in their analysis---in Ref.~\cite{Asokan:2022usm} it is explained,
why the higher one was overlooked.} instead of a single one
(for more refined analyses of the finite momentum lattice data see Refs.~\cite{Guo:2018gyd,Guo:2018kno,Guo:2018tjx,Lutz:2022enz}).
Moreover, when chirally extrapolated to the physical pion mass the two poles were located at
2105 and 2451~MeV, respectively---in line with the earlier findings of Ref.~\cite{Guo:2006fu}. The reason can be read off, e.g., Fig.~\ref{fig:Dpispectrum}
showing the results of Ref.~\cite{Du:2017zvv}: A naive Breit-Wigner
fit of the data will naturally return a mass at 2.3~GeV. However, in the molecular picture
this structure appears as the interplay of the mentioned two poles, one above and one below 2.3~GeV. Moreover,
chiral symmetry constraints
pull the pole location to lower values~\cite{Du:2019oki} already in the absence
of a second pole. This is backed by an analysis of the phase of the
$\pi D$ amplitude also extracted in Ref.~\cite{LHCb:2016lxy} which is found inconsistent
with a Breit-Wigner pole at $2.3$~GeV, but fully consistent with the amplitude of Ref.~\cite{Liu:2012zya}
as demonstrated in Refs.~\cite{Du:2020pui}. Thus, the SU(3)-flavor violation within the  multiplets of the molecular
states is characterised by
the quark mass dependence 
of the pseudo-Goldstone bosons, which is build into the energy dependence of Eq.~(\ref{eq:LOhl}), while
group theory decides, which multiplets exist and their hirarchy---for a more general discussion of two-pole structures see Ref.~\cite{Meissner:2020khl}.

Finally we would like to comment why existing experimental and lattice data are in striking conflict
with both a quark--anti-quark structure of the states as well as with a diquark--anti-diquark structure
for the states discussed above.
To exclude the former it is sufficient to fully establish the existence of the states in the flavor
$[\mathbf{6}]$---this was done e.g. in Refs.~\cite{Gregory:2021rgy,Yeo:2024chk} via lattice simulations
with pion masses high enough that even this state gets nearly bound as predicted in Ref.~\cite{Du:2017zvv}.
The striking difference between two-hadron structures and the diquark--anti-diquark picture is that in the
former there are no states in the $[{\mathbf{\overline{15}}}]$ representation, while in the latter they should 
be there, if both axial-vector and scalar diquarks exist as relevant degrees of freedom~\cite{Guo:2025imr}---in Ref.~\cite{Maiani:2024quj}
a phenomenology consistent with that of the two-hadron states is found, however, at the cost of
abandoning the axial-vector diquarks that on the other hand are needed for a consistent diquark phenomenology
in other systems~\cite{Barabanov:2020jvn}.
The key effect of states in the $[{\mathbf{\overline{15}}}]$ representation that would emerge in the 
diquark--anti-diquark picture is that low the lying $D$-meson spectra in the $0^+$ and $1^+$ sector should be very 
different, while they would be very similar in the molecular scenario.
The reason for the pattern for the compact tetrquarks is that the mass difference
of about 200~MeV
between the scalar and the axial-vector diquark is roughly the same 
for heavy-light and light-light diquarks. Accordingly are the states in the $[{\mathbf{\overline{15}}}]$ representation
in the scalar sector about 400 MeV heavier than the state in the $[\mathbf{\bar 3}]$ and in the axial-vector
sector about equally light, simply since scalar (axial-vector) quantum numbers are reached by either
combining two scalar or two axial-vector diquarks (a scalar and an axial-vector diquark). 
Since the lattice simulations of Ref.~\cite{Gregory:2025ium} in a flavor symmetric setting  
show consistent energy levels for both quantum numbers, the diquark picture is excluded for the
lightest positive parity open charm states.

\section{Doubly heavy states: $X(3872)$ and its isovector partner $W_{c1}$}

\begin{figure}
\includegraphics[width=\linewidth]{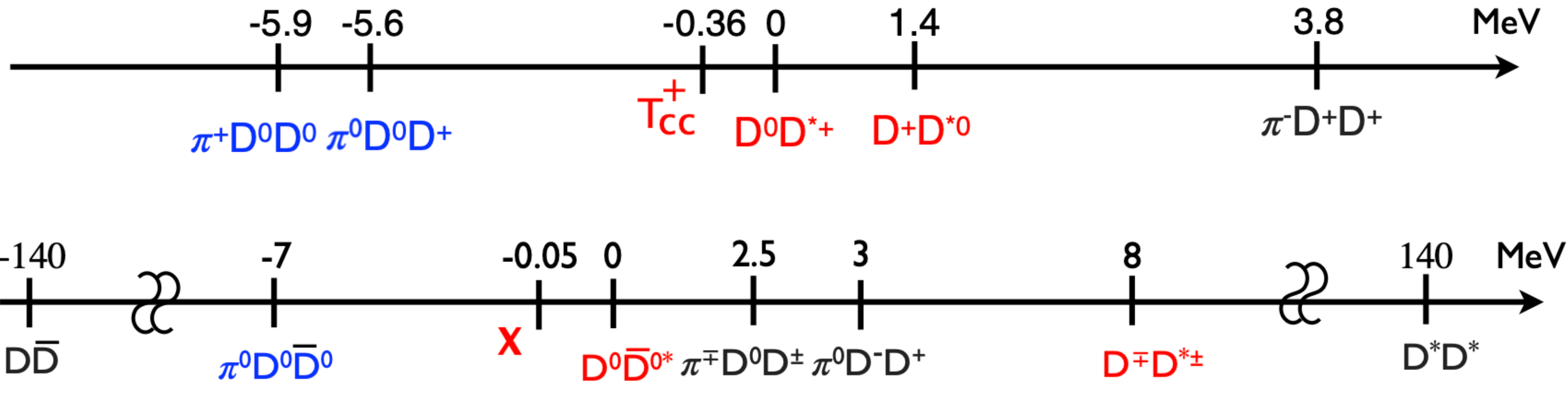}
\caption{The upper and the lower plot shows relevant scales for the $T_{cc}(3875)$ and the $X(3872)$, respectively. Figs. courtesy of V. Baru.
 \label{fig:TccXscales}
}
\end{figure}

The first works studying the $X(3872)$ and the $T_{cc}(3875)$  with non-perturbative pions are Ref.~\cite{Baru:2011rs}
and Ref.~\cite{Du:2021zzh}, which also form the basis of
the results presented here---alternatively Ref.~\cite{AlFiky:2005jd} propose to work without pions and Ref.~\cite{Fleming:2007rp} to include
them perturbatively, but the energy range of applicability of those approaches is more limited and the effects
of isospin violation cannot be captured completely, as is illustrated by Fig.~\ref{fig:TccXscales}.

The $T_{cc}(3875)$, discovered by LHCb~\cite{LHCb:2021vvq}, is seen as a prominent structure in the $D^0D^0\pi^+$
invariant mass distribution resulting from prompt $pp$ collisions---it thus contains two charm quarks
and has no strong decay channel besides two $\pi DD$ channels (see upper plot of Fig.~\ref{fig:TccXscales}). On the one hand 
this makes it an ideal playground for theoretical studies both within EFT as well as lattice QCD, on the other hand
it calls for a careful treatment of three-body effects, which need  both an energy dependent $D^*$ width as well
as a full treatment of the pion propagator including its energy dependence and non-locality induced by the
$D$-meson recoil terms (for a detailed discussion of this effect see Sec. 3 in Ref.~\cite{Lensky:2005hb}).
This is especially necessary to ensure that
the imaginary parts that the two contributions induce show the onset of a three-body phase space and that
 the intermediate two-$D$ state
that is present while the pion is in flight has the proper Bose-symmetry.
As a result of this the imaginary part of the pole location is very sensitive to the
formalism: the pole location that emerges from the full calculation of Ref.~\cite{Du:2021zzh} is $(-356\pm 39-i(28\pm 1))$~keV, 
while an omission of the cut in the potential changes the imaginary part to $-(18\pm 1)$~keV~\cite{Du:2021zzh}.
The effective range extracted from the full calculation, corrected for isospin violation following
Ref.~\cite{Baru:2021ldu}, is $r_0=(1.38\pm 0.85)$~fm---thus the $T_{cc}$--phenomenology is fully consistent with
its being a $DD^*$ hadronic molecule.
Moreover, already for pion masses only slightly above the physical value the $D^*$ gets stable, such that the pion
exchange induces a very close by left-hand branch point~\cite{Du:2023hlu} and, through this,
the chiral extrapolation gets
very non-trivial, employing the chiral effective field theory described above~\cite{Abolnikov:2024key}. 
A confirmation of this behavior from careful lattice QCD studies for very near physical pion masses
should provide further strong evidence for a molecular nature of $T_{cc}(3875)$.

\begin{figure}
\includegraphics[width=\linewidth]{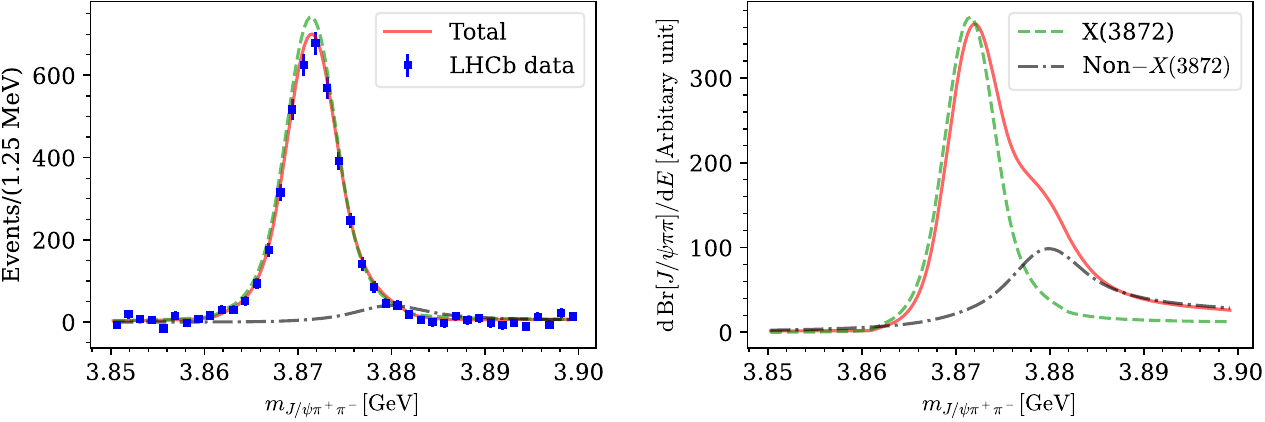}
\caption{Results for the $J/\psi \pi\pi$ spectra of Ref.~\cite{Ji:2025hjw}. Left panel:  Fit result data extracted from
 $B^+\to K^+ J/\psi \pi\pi$~\cite{LHCb:2022jez}; right panel: Estimate for the spectrum from $B^0\to K^0 J/\psi \pi\pi$.
 In both cases the theory curves are convolved with the LHCb resolution. The red solid line shows the full result,
 the green dashed one just the contribution from the $X(3872)$ and the black dot-dashed line,
 featuring the $W_{c1}^0$ signal, the difference of the two (subtracted on the 
 amplitude level). The $W_{c1}^0$ signal is enhanced (reduced) in the $B^0$  ($B^+$) decay, since here the charged $D\bar D^*$ states
 are more frequently (less frequently) produced than the neutral one.
 \label{fig:Jpsipipispectra}
}
\end{figure}

Finally we report briefly the most recent and most precise study of the $X(3872)$ reported in Ref.~\cite{Ji:2025hjw}.
Also in this work the full three-body dynamics as well as the leading isospin violating effects are taken care off,
with the latter being mass differences within isospin multiplets of the participating particles
(pions and $D$-mesons) as well as $\rho{-}\omega$ mixing. A fit
to the data for $B^+\to K^+(J/\psi\pi^+\pi^-)$ by LHCb~\cite{LHCb:2020fvo,LHCb:2022jez} and  
$e^+e^-\to \gamma (D^0\bar{D}^{0}\pi^0/J/\psi\pi^+\pi^-)$
by BESIII~\cite{BESIII:2023hml} confirmed the claim of Ref.~\cite{Zhang:2024fxy} that in addition to the mostly
isocalar $X(3872)$ there needs to be a predominantly isovector state, dubbed $W_{c1}$, present. The pole
of the $X(3872)$ 
is found to be a {quasi-}bound state with a significance of $2.8\,\sigma$ at 
$(-160^{+57}_{-74}-125^{+23}_{-38}\,i) \rm keV$, relative to the nominal
$D^0\bar{D}^{*0}$ threshold and the $W_{c1}^0$ pole is found at $(3.1\pm0.7+ 1.3^{+1.9}_{-0.6}\,i)\ \rm MeV$ relative to the $D^+ D^{*-}$ threshold on a so called hidden Riemann sheet---see below for a more detailed discussion. In 
the existing high statistics data the latter state appears merely as a slight modulation of the
the line shape provided by the $X(3872)$ alone, however,  as a very non-trivial prediction it is 
demonstrated in Ref.~\cite{Ji:2025hjw} that this signal should be significantly more prominent
in $B^0\to K^0\left(\left.D^0\bar{D}^{0}\pi^0\right/J/\psi\pi^+\pi^-\right)$, which can be studied at both Belle-II and LHCb.
The origin of the difference is a combination of the expectation that in $B^0\to K^0(X/W_{c1}^0)$ decays, contrary to $B^+\to K^+(X/W_{c1}^0)$,
 the charged intermediate states, $D^\pm D^{*\, \mp}$,
are more frequently produced than the neutral ones and
that the $X(3872)$ and the $W_{c1}^0$ sit rather close to the neutral and charged threshold, respectively.
Fig.~\ref{fig:Jpsipipispectra} shows the fit to the existing LHCb data for 
$B^+\to K^+J/\psi\pi^+\pi^-$ (left panel), where the fit returned for the ratio of couplings of the source to
the charged channel over that to the neutral one 0.5, as well as the prediction for $B^0\to K^0J/\psi\pi^+\pi^-$ (right panel),
where the same ratio is estimated to be 2. In the figure the red solid line shows the full result, while the
green dashed the one of the $X(3872)$ only, which is given by a Flatte parametrisation 
with parameters fixed to reproduce the pole location as well as the pole residues as suggested 
in Ref.~\cite{Heuser:2024biq}. For the black dot-dashed curve the $X(3872)$ contribution was subtracted from the full
result on the amplitude level---it thus shows the $W_{c1}$ signal.

To better understand the phenomenology of the $X$ and the $W_{c1}$---and in particular why the $W_{c1}$ was not seen
prominently in data yet---and to see why it strongly points at a molecular nature of both states,
we need to discuss in more detail the pole locations of $W_{c1}^\pm$, $W_{c1}^0$ and $X$,
respectively, largely following Ref.~\cite{Zhang:2024fxy}. The natural explanation for the charged $W_{c1}$ states
not producing a prominent signal in data is, that they are virtual states---those are below threshold poles on the
unphysical Riemann sheet, labeled $R_-$, where the subscript denotes the sign of the imaginary part 
of the two-hadron momentum. Virtual states, which are necessarily 
of molecular nature~\cite{Matuschek:2020gqe}, can leave a significant imprint in observables only, if they
are located very close to the threshold (see, e.g., the neutron-neutron system, which has a virtual pole only 66~keV below threshold, 
leading to a scattering length of 18.5~fm). The $W_{c1}^\pm$ poles are found
$8^{+8}_{-5}$~MeV below the $D^0D^{*\, \pm}$  threshold\footnote{For simplicity in this discussion we
neglect the effects from the $D^*$ decay and other inelastic channels.}, such that their impact on the respective lineshapes
is small.
In particular, in comparison with $X(3872)$, which shows its full peak below threshold 
as a quasi-bound state, the peak height prior to convolution with the energy resolution is orders
of magnitude smaller~\cite{Zhang:2024fxy}.
 If there were no isospin violating effects (for charged and neutral
$D^{(*)}$-mesons being mass degenerate and an isospin-symmetric interaction),
 the $W_{c1}^0$ would be at the same location. 
If we assume in addition that the isovector and isoscalar interactions are of equal strength (note that
this assumption allows for a successful phenomenology for light and medium heavy
nuclei~\cite{Lu:2018bat}), 
also the pole of the $X$ appears as
a virtual state. Thus the system hosts two poles, one close to the  neutral $D\bar D^*$ threshold, one
close to the charged one---the transitions between the two channels, which scale
with the difference in isoscalar and isovector interaction strength, are at this point
assumed negligible; an assumption to be lifted below.
 Introducing isospin violation (most importantly by the mass differences of the neutral and the charged $D$-mesons) 
 affects the charged $W_{c1}$
states only mildly, however, in the neutral sector the thresholds move apart and the two poles start 
to feel each other---the system becomes a coupled channel problem with, already for infinitesimal
channel coupling, the two poles sitting on sheets $R_{-+}$ and $R_{+-}$,
respectively. This channel-coupling has two important effects: On the one hand the two poles repel each other---the now
lower lying pole feels more attraction, eventually moving from the $R_{-+}$ sheet to the
$R_{++}$ sheet, explaining naturally why the
physical $X(3872)$ appears as bound state in data. Note that this trajectory is also possible only,
when the $X$ is a molecular state, since it starts as a virtual state and needs a strong channel coupling.
This claim is consistent with a direct evaluation of the compositness of the $X$ using
the method of Ref.~\cite{Li:2021cue} (a generalisation of the Weinberg criterion
to include range corrections), which gives a probability of $97\pm 2$ \% to find the
two-hadron component in the $X$ wave function. At the same time
 the higher lying pole, identified with the $W_{c1}^0$, gets pushed up and acquires a width,
 since it can now decay into the lower lying neutral $D\bar D^*$ channel. In other words: the $W_{c1}^0$
 pole moves into the complex plane. However, since it was sitting on the $R_{+-}$ sheet, it stays there
 with no direct connection to the physical axis besides exactly at the charged $D\bar D^*$ threshold:
 it sits on a hidden sheet. The scenario
 outlined here agrees to that of scenario V4 of Ref.~\cite{Zhang:2024qkg}. 
That the lower lying $X$ state is predominantly isoscalar and the higher lying $W_{c1}$ is predominantly
isovector follows most probably from the observation that the isocalar interaction is in reality more attractive 
than its isovector counter
part~\cite{Ji:2025hjw}.  

\section{Summary}

These are clearly exciting times for the spectroscopy of single- and double-heavy states,
since high quality data meet sophisticated theoretical tools, namely lattice QCD and effective
field theories. Here we focussed on two sectors, namely the lowest lying positive parity open charm states
and the states sitting close to the $D\bar D^*$ and $DD^*$ thresholds, respectively.
The effective field theory ideal to study hadronic molecules is unitarised chiral perturbation theory,
which was employed here in all cases.
In the first part of the presentation we demonstrated that the phenomenology and the
lattice QCD data of the open
charm states in the $0^+$ and $1^+$ sectors not only are fully consistent with what is expected from a two-hadron
nature of the states, but also that they are inconsistent with a $c\bar q$ as well as diquark--anti-diquark
structure. In the second part, where the doubly heavy states with $J^{PC}=1^{++}$~\footnote{Clearly,
for the isovector states the $C$ assignment can be applied to the neutral state only.} were discussed,
evidence was presented that there exits an isovector partner of the $X(3872)$ whose existence can
be confirmed by studies of $B^0\to K^0(X/W_{c1}^0)$. Moreover, also for these states compelling
evidence is presented that they can be considered molecular with the same level of rigour as the deuteron.

\vspace{0.2cm}

{\it Acknowledgments}---The author thanks the organisers of Hadron2025 conference for the excellent organisation of a
very educating conference, Vadim Baru, Xiang-Kun Dong, Menglin Du, Evgeny Epelbaum, Feng-Kun Guo,
Teng Ji, Ulf-G. Mei\ss ner, Alexey Nefediev and many others for the pleasant and productive collaboration on the work presented, 
Feng-Kun Guo and Ulf-G. Mei\ss ner for valuable comments on the manuscript,
 and the CAS President's International Fellowship Initiative (PIFI) under
 Grant No.\ 2025PD0087 for partial support. Part of the work was supported by the MKW NRW
under funding code NW21-024-A.

%\bibliographystyle{elsarticle-num} 
%\bibliography{reference}

%\end{document}

\end{document}